# Best estimator for bivariate Poisson regression

André GILLIBERT[ab]*†, Jacques BÉNICHOU[bc] and Bruno FALISSARD[a]

[a] INSERM UMR 1178, Université Paris Sud, Maison de Solenn, Paris, France.

[b] Department of Biostatistics and Clinical Research, CHU Rouen, Rouen, F 76031, France

[c] Inserm U 1181, Normandie University, Rouen, France

* Correspondence to: André GILLIBERT, Department of Biostatistics and Clinical Research, CHU Rouen, Rouen, F 76031, France

†E-mail: andre.gillibert@chu-rouen.fr

# Abstract

**INTRODUCTION**: Wald's, the likelihood ratio (LR) and Rao's score tests and their corresponding confidence intervals (CIs), are the three most common estimators of parameters of Generalized Linear Models. On finite samples, these estimators are biased. The objective of this work is to analyze the coverage errors of the CI estimators in small samples for the log-Poisson model (i.e. estimation of incidence rate ratio) with innovative evaluation criteria, taking in account the overestimation/underestimation unbalance of coverage errors and the variable inclusion rate and follow-up in epidemiological studies.

**METHODS**: Exact calculations equivalent to Monte Carlo simulations with an infinite number of simulations have been used. Underestimation errors (due to the upper bound of the CI) and overestimation coverage errors (due to the lower bound of the CI) have been split. The level of confidence has been analyzed from 0.95 to $1\text{-}10^{-6}$, allowing the interpretation of P-values below $10^{-6}$ for hypothesis tests.

**RESULTS**: The LR bias was small (actual coverage errors less than 1.5 times the nominal errors) when the expected number of events in both groups was above 1, even when unbalanced (e.g. 10 events in one group *vs* 1 in the other). For 95% CI, Wald's and the Score estimators showed high bias even when the number of events was large (≥20 in both groups) when groups were unbalanced. For small P-values ($<10^{-6}$), the LR kept acceptable bias while Wald's and the score P-values had severely inflated errors (×100).

**CONCLUSION**: The LR test and LR CI should be used.

# 1 Introduction

Generalized Linear Models (GLMs) are a family of statistical models, including the widely used logistic regressions and Poisson regressions. Several hypothesis tests and estimators of CIs (CIs), exist for parameters of these models. The best known tests are Rao's score test (also known as the score test or Lagrange multiplier Test), Wald's test and the generalized likelihood ratio test (GLRT) [1]. These three tests are equivalent when there is no nuisance parameter [2], *i.e.* a parameter such as the base incidence rate (*e.g.* incidence rate of group 1) for estimating an incidence rate ratio: it is not the parameter that we wish to estimate, but our estimation of the incidence rate ratio depends on it, so that uncertainty of this nuisance parameters generates an error or a bias in the estimation of the parameter that we wish to estimate. Since nuisance parameters stabilize when the sample size grows, the three tests are asymptotically equivalent for GLMs. The statistics of Rao's score test [3] and of the GLRT (after transformation) [4] have both an asymptotic chi-square distribution. Wald's statistic, for GLMs, is asymptotically normally distributed [5]. By test inversion, CIs estimators can be built from these three hypothesis tests.

> Note that the generalized likelihood ratio test (GLRT) is often abbreviated to likelihood ratio test (LRT), but is actually not equivalent. The GLRT statistic is the ratio of the likelihood of the observed data under the null hypothesis and the maximum likelihood (ML) estimate over the complete space of the tested parameter while the LRT defined in Neyman-Pearson's lemma is based on the ratio of likelihood for two fully specified parameters $\theta_0$ and $\theta_1$ [6]. The GLRT can be used to reject a hypothesis when all values in a parameter space are possible while the LRT relies on a parameter space restricted to two values, is the latter being uncommon; it can help choose between two hypotheses when no other hypothesis is possible. In the rest of this document, the term likelihood ratio (LR) is considered synonymous to the generalized likelihood ratio since the GLRT is, by far, the most used method and is usually called LR.

Statistical software do not rely on the same default estimator (e.g. likelihood ratio CI for R, Wald's CI for SAS, SPSS, Stata) and may even be inconsistent: for instance, by default R (version 3.6.2) uses Wald's P-values and likelihood ratio CIs for GLMs.

On finite samples, some estimators and tests may behave better than others. The standard estimators (Score, Wald, LR) are supplemented by more anecdotal estimators such as the penalized likelihood estimators (Firth [7], Kenne [8]) and Hirji's exact estimator [9]. Since these estimators have been designed to reduce bias, we will analyze their behavior as well.

A quick review of articles (articles reviewed by one researcher, with their supplementary material if available) published between September and November 2018 in the British Medical Journal, The Lancet, The Journal of the American Medical Association, The New England Journal of Medicine, and the Annals of Internal Medicine, showed that, out of 203 research articles, 198 were non-meta-analysis articles of which 58 used a logistic regression (fixed effects, conditional, mixed effects or GEE) for primary, secondary or post hoc analysis, of which 46 tested hypothesis or provided CIs of odds ratio (12 logistic regressions were used for multiple imputation, propensity score matching or were an intermediate step for a more complex calculation). Out of the 46 original research articles using a logistic regression with a hypothesis test or CI on an odds ratio, 9 (20%) gave some information about the estimator or test used and 6 (13%) [10–15] gave explicitly an estimator or test name while the other 3 (7%) [16–18] gave indirectly the estimator, by sharing the script or software package used. Of the 6 articles with explicit estimator or test sharing, one of them [11] added the specification of the test after retraction and replacement of the article due to errors in statistical analyses. Therefore, most articles do not specify the estimator used.

The Poisson distribution is asymptotically equivalent to the Binomial distribution if the proportion tends towards zero. The Gaussian approximation of the binomial distribution is best when the actual proportion is close to 50%, leading to small bias in logistic regressions. For low proportions and high proportions, the binomial distribution is highly skewed and the Gaussian approximation is poor. For a logistic regression, the two extreme scenarii are equivalent since a high proportion can be transformed to a low proportion by inverting the outcome (analyzing non-events rather than events). Therefore, we concentrate our analysis in the scenario of low proportion. A logit-binomial and log-Poisson model are equivalent when the denominator tends towards infinity. Therefore, we considered that the log-Poisson model is the worst case scenario (most biased for Wald's CI) of the logistic regression. That is why we analyzed the Poisson regression only.

The objective of this work was to assess the coverage bias of bivariate Poisson regression CI estimators with innovative evaluation criteria in order to define their validity conditions and provide useful advice on which estimator to use and when to use it.

# 2 Rationale and Methods

## 2.1 Scenario analyzed

The assessment of the ratio of two Poisson distribution has been analyzed. We note $Y_1 \sim P(\lambda_1)$ and $Y_2 \sim P(\lambda_2)$ two independent variables following Poisson distributions and $r_1$ and $r_2$ the offset (e.g. number of subjects-years of follow-up). The ratio $\rho = \frac{\lambda_2/r_2}{\lambda_1/r_1}$ can be interpreted as an incidence rate ratio. The estimation of this IRR is equivalent to the estimation of $\tau = \frac{\lambda_2}{\lambda_1}$ after a change of variable. Indeed, comparing $\rho$ to a value $\rho_0$ in a GLM with offset is equivalent to comparing $\tau$ to a value $\rho_0 \times \frac{r_2}{r_1}$ in a GLM without offset. That is why we only analyzed the scenario without offset. The model can be written:

$$\log(E[Y_1]) = \beta_0$$

$$\log(E[Y_2]) = \beta_0 + \beta_1$$

With $J = \exp(\beta_1)$ the IRR that we wish to estimate.

The plane of all possible combinations of $(\lambda_1, \lambda_2)$ will be analyzed from $\lambda_1 = 0.5$ and $\lambda_2 = 0.5$ to $\lambda_1 = 104$ and $\lambda_2 = 104$ by increments of 0.05.

## 2.2 Evaluation criteria: one-sided unconditional risks

The coverage bias of two-sided 95% CIs is usually assessed by the coverage error defined as the actual coverage minus the nominal coverage. For the realization of a CI, a coverage fault can be an overestimation (lower boundary of the CI above the actual value) or an underestimation (upper boundary of the CI below the actual value). The overall coverage error is computed by adding the probability of overestimation to that of underestimation; doing so, they are somehow assumed to be equivalent. Underestimating an incidence rate ratio in 4.9% of cases and overestimating it in 0.1% of cases will be seen as an unbiased CI estimator by this statistic, since the overall coverage will be 100% − 4.9% − 0.1%=95%, but will falsely reassure when answering to the research question "may this exposition raise the incidence of that disease?". Therefore, we think that, in most cases a two-sided CI should behave as the intersection of two one-sided CIs with the same risk of non-coverage.

That is why we will analyze separately the non-coverage due to the lower and upper boundaries of the CI estimators. A balanced CI estimators is wished.

We define $\alpha_L = \alpha_U = \alpha$ the nominal probabilities that a two-sided CI be strictly above or strictly below the actual statistic $\rho = \lambda_2/\lambda_1$. This definition seeks for balanced CIs. For a two-sided 95% CI, $\alpha = 0.025$.

Assuming that $\lambda_1$ and $\lambda_2$ are fixed, we respectively define the ***conditional risks*** as $\alpha'_L$ and $\alpha'_U$ as the actual probabilities that the CI estimator is strictly above or strictly below the actual statistic $I$. That is, if $Y_1 \sim P(\lambda_1)$ and $Y_2 \sim P(\lambda_2)$ are two random variables following Poisson distributions, and $C(y_1, y_2) = [L(y_1, y_2); U(y_1, y_2)]$ is a CI estimator with a lower boundary $L$ and a upper boundary $U$, computed from the observed number of events $y_1$ and $y_2$ , then $\alpha'_L = \Pr\left(L(Y_1, Y_2) > \frac{\lambda_2}{\lambda_1}\right)$ and $\alpha'_U = \Pr\left(U(Y_1, Y_2) < \frac{\lambda_2}{\lambda_1}\right)$.

In studies estimating incidence rate ratios, $\lambda_1$ and $\lambda_2$ depend on the duration of follow-up and inclusion rate. The inclusion rate is rarely controlled and can be considered random and the follow-up is not always perfectly controlled. Therefore, we can consider that $r_1$ and $r_2$ are random variables. Although the incidence rate ratios $I_1 = \frac{\lambda_1}{r_1}$ and $I_2 = \frac{\lambda_2}{r_2}$ may be constant, the randomness of $r_1$ and $r_2$ leads to random $\lambda_1 = I_1 \times r_1$ and $\lambda_2 = I_2 \times r_2$. We note $\Lambda_1$ and $\Lambda_2$ the random variables representing the expected number of events, assuming they follow log-normal distributions with a geometrical standard deviation equal to 1.10 and expectancies respectively equal to $\lambda_1$ and $\lambda_2$. The random distribution of $\Lambda_1$ and $\Lambda_2$ were chosen arbitrarily. In sensitivity analyzes, the geometrical standard deviation 1.10 was changed. We define random variables $Y''_1$ and $Y''_2$ built from a two-steps procedure: realizing $\Lambda_1$ and $\Lambda_2$ as $l_1$ and $l_2$, then, drawing $y''_1$ and $y''_2$ from Poisson distributions with expectancy $l_1$ and $l_2$. In this two-steps experiment, $l_1$ and $l_2$ represents the expected number of events in both groups after the total duration of follow-up ($r_1$ and $r_2$) are realized and before the actual number of event occur. The values $y''_1$ and $y''_2$ represent the observed number of events in both groups at the end of the experiment.

We define the ***unconditional risks*** $\alpha''_L$ and $\alpha''_U$ as $\alpha''_L = \Pr\left(L(Y''_1, Y''_2) > \frac{\Lambda_2}{\Lambda_1}\right)$ the actual probability that the confidence interval be strictly above the actual incidence rate ratio and $\alpha''_U = \Pr\left(U(Y''_1, Y''_2) < \frac{\Lambda_2}{\Lambda_1}\right)$ the probability that the confidence interval be strictly below the actual incidence rate ratio.

The nominal coverage will be 95% for the primary analysis, i.e. $\alpha = \alpha_L = \alpha_U = 0.025$. There is no consensus limit of what is an "acceptable" coverage bias. The limit has been arbitrarily set to $\frac{\alpha}{1.5} = 0.01667$ and $\alpha \times 1.5 = 0.0375$. Less strict (1.25) and more strict (2 and 10) multiplicative thresholds are shown in figures.

## 2.3 Evaluation criteria: interval relative half-width

As the coverage errors are separately assessed for both boundaries of the CI, the width of the CI is better split in two halves and assessed separately: the lower and upper half-widths are equal to the distance between ML point estimate (even for Firth's and Kenne's CI) and, respectively, the lower and upper boundaries of the CI.

The half-width is highly dependent on the number of events observed. For instance, for $y_1 = 1$ and $y_2 = 10$ the upper half-width of the ML LR 95% CI and Hirji's mid-P 95% CI are respectively equal to 173.4 and 210.1 (ratio of the two half-widths = 0.83), but for $y_1 = 100$ and $y_2 = 100$ the half-widths are respectively equal to 0.3200 and 0.3206 (ratio = 0.9982). In these two scenarii, the half-widths are different by order of magnitudes (~200 *vs* ~0.3) but the ratio of half widths are much closer (0.83 *vs* 0.9982). Therefore, a figure showing crude half-widths for a large range of values would be illegible but a figure showing ratios of half-widths of a CI to a reference CI is much more legible. Therefore, we displayed ratios of half-widths of estimators relative to each other.

## 2.4 Estimators

The following CI estimators of bivariate log-Poisson regressions have been analyzed:

1) The CI constructed by inversion of Rao's score test (score CI)
2) Wald's asymptotic normal CI (Wald's CI)
3) The profile likelihood ratio CI based on ML estimates (ML LR CI)
4) Wald's CI in a GLM fitted by Firth's penalized likelihood ratio estimator [7] (Wald-Firth CI)
5) Non-Penalized LR CI in a GLM where the point estimates are estimated by Firth's penalized LR estimator (LR1-Firth CI)
6) Penalized LR CI in a GLM where the point estimate is estimated by Firth's penalized LR estimator (LR2-Firth CI)
7) Wald's CI [8] in a GLM fitted by Kenne's penalized LR estimator (Wald-Kenne CI)
8) Hirji's exact estimator of the GLM CI [9] (Hirji's CI)
9) Hirji's exact estimator of the GLM CI [9] with mid-P value modification [19] (Hirji mid-P CI)

All these CI estimators can be constructed by inversion of hypothesis tests. Consequently, the results shown for CIs can be extrapolated to corresponding hypothesis tests.

When the number of events $y_1$ or $y_2$ is equal to zero, some estimators do not converge (Rao's score, Wald's CI, the ML LR CI). Penalized LR estimators (Firth, Kenne) and exact estimators do converge unless both $y_i$ are zero. For consistency, it has been assumed that whenever one or the other $y_i$ is zero, no CI will be computed. Therefore, all results of this work are conditional to the fact that both $y_i$ are strictly positive.

## 2.5 Computation method of CI

### 2.5.1 The Wald, ML LR and score CI

As these estimators are conditional to $y_1 + y_2$, they have been shown to be equivalent to the estimation of a binomial proportion in a univariate (intercept-only) logistic regression after a change of variable. Applying the function $p \to \frac{p}{1-p}$ to both boundaries of the CI of the binomial distribution yields the bivariate Poisson regression CI. This gave simple analytical solutions to Wald's and the score CI. The ML LR CI is harder to compute and required numerical inversion of the family of function $r(x) = a \times \log(x) + b \times \log(x+1) + c$ where $a$, $b$ and $c$ are constant parameters of the function. After a change of variable $y = \log(x)$, the Newton-Raphson algorithm converged in three iterations to more than 10 decimal places (data not shown).

### 2.5.2 Firth's LR CI

The LR CI has been computed by inverting a hypothesis test. The Newton-Raphson algorithm with numerical derivation was used. The hypothesis test has been performed by fitting the model without constraining the slope (alternative hypothesis, 2 degrees of freedom) and with a fixed slope (null hypothesis, 1 degree of freedom). The fitting process of both models was performed by Firth's penalized likelihood. Then, either the log-likelihood (LL) or the penalized log-likelihood (PLL) of both fitted models was computed, respectively for LR1-Firth and LR2-Firth CIs. Twice the difference of LL or PLL, was approximated to a chi square distribution at one degree of freedom under the null hypothesis.

### 2.5.3 Hirji's exact CI

Hirji described his estimator for the logit-binomial regression [9] but it can be generalized to the Poisson distribution. Indeed, a Poisson distribution is equivalent to a binomial distribution with a sample size (parameter $n$) tending towards infinity. Like the Wald, ML LR and score CI, Hirji's exact CI is equivalent to the binomial exact CI. It can be based on the Clopper-Pearson CI with or without mid-P modification.

## 2.6 Computation method of non-coverage probabilities

### 2.6.1 Conditional risks

Coverage errors, defined as one-sided risks, as seen in section 2.2, are dependent on $\lambda_1$ and $\lambda_2$ the expected number of events (*e.g.* crude incidence of events) but not on $r_1$ and $r_2$, the number of patients-years of follow-up. Therefore, all our calculations are only based on $\lambda_1$ and $\lambda_2$, and the set of possible values for $\lambda_1$ and $\lambda_2$ is explored up to $\lambda_1 = 104$ and $\lambda_2 = 104$ by increment of 0.05, leading to a large matrix of coverage errors, graphically represented as a colored surface.

Rather than based on Monte Carlo sampling, coverage errors have been computed by almost exact probability calculation. This made it possible to quickly estimate coverage errors for a confidence level equal to $1 - 10^{-6}$. The algorithm for the lower bound risk $\alpha'_L$ conditional to Poisson parameters $\lambda_1$ and $\lambda_2$ is described below:

1) Input of the algorithm : $\lambda_1$ and $\lambda_2$, the expected number of events, and $\alpha_L$ the nominal one-sided coverage error
2) Compute quantiles $10^{-9}$ and $1 - 10^{-9}$ for both $P(\lambda_1)$ and $P(\lambda_2)$ distributions
3) Sets the overall sum to zero
4) For all $y_1$ and $y_2$ (target observation) between the respective quantiles of Poisson distributions (matrix of all possible pairs of observations $(y_1, y_2)$)
    a. Compute the CI for $y_1$ and $y_2$
    b. If the CI is entirely above $\lambda_2/\lambda_1$, adds the probability that $Y_1 = y_1$ and $Y_2 = y_2$ (product of both probabilities) to the overall sum, with $Y_1 \sim P(\lambda_1)$ and $Y_2 \sim P(\lambda_2)$

At the end of the loop, the overall sum is the probability that the CI is entirely above the actual ratio $\lambda_2/\lambda_1$. The same algorithm applies to $\alpha'_U$ but the upper boundary is compared to $\lambda_2/\lambda_1$. A small adjustment of the matrix of probabilities has been applied to make sure that the sum of all probabilities of $(y_1, y_2)$ pairs is equal to 1, rather than $(1 - 10^{-9})^2$ as would be expected from the quantiles of the Poisson distribution described at step 1 of the algorithm. This algorithm is equivalent to Monte Carlo sampling with an infinite number of simulations, with an error less than $10^{-9}$.

Conditional risks have been computed for all $\lambda_1$ and $\lambda_2$, from 0 to 40 and from 100 to 104 by increment of 0.05. This defines the size of figures pixels.

### 2.6.2 Unconditional risks

The log-normal distributions of $\Lambda_1$ and $\Lambda_2$ define a bivariate distribution. The bivariate distribution, assuming $\Lambda_1$ and $\Lambda_2$ are independent, was approximated to a bivariate discrete distribution with a 20×20 grid (40×40 for the sensitivity analysis with geometrical standard deviation = 1.20). The support of this discrete distribution was calculated to align on the grid (precision 0.05) that had been used for the computation of conditional risks. Then, the probabilities of each $(l_1, l_2)$ pair was multiplied by the conditional $\alpha'_L$ or $\alpha'_U$ risk, then summed up to estimate the unconditional $\alpha''_L$ and $\alpha''_U$ risks.

# 3 Sensitivity analyzes

For unconditional risks, the geometrical standard deviation has been set at 1.10 in the primary analysis and changed to 1.05 and 1.20 in sensitivity analyzes.

# 4 Secondary analyzes

The two-sided confidence levels has been set at 0.95 in the primary analysis and changed to 0.80, 0.90, 0.999 and $1{-}10^{-6}$ in secondary analyzes.

# 5 Results

## 5.1 Unconditional coverage

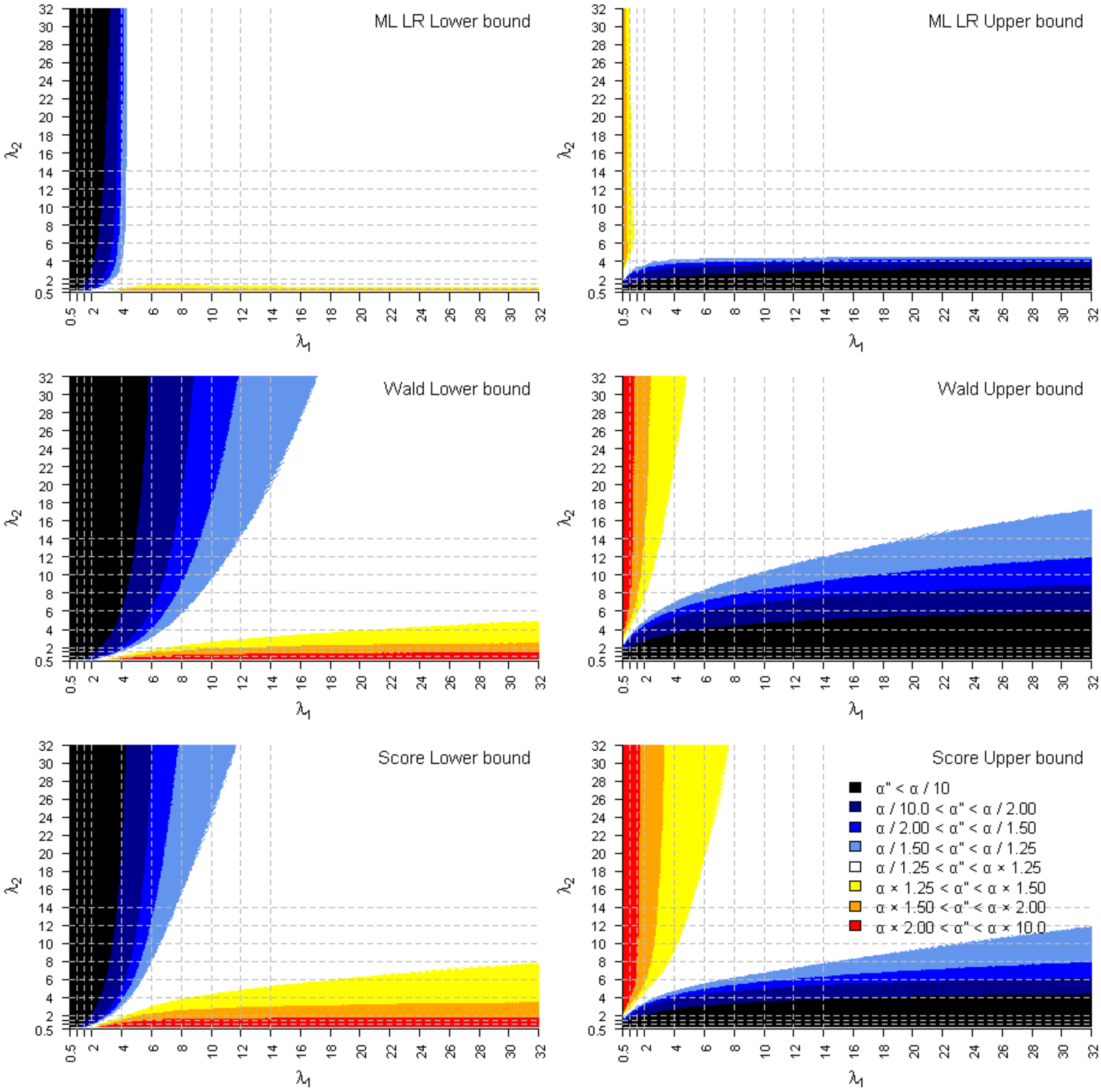

*Figure 1 : unconditional $\alpha''_L$ and $\alpha''_U$ risks for the 95% two-sided likelihood ratio confidence interval (ML LR), 95% two-sided Wald confidence interval and 95% two-sided Score confidence interval of a ratio $\Lambda_2/\Lambda_1$ of two Poisson variables, according*

*to $\lambda_1$ (x axis) and $\lambda_2$ (y axis) the expectancies of the Poisson distribution parameters assuming these parameters follow a log-normal distribution having a geometrical standard deviation equal to 1.1. The white, yellow and light blue zones show the zones of tolerated risk deflation/inflation by a factor less than 1.5.*

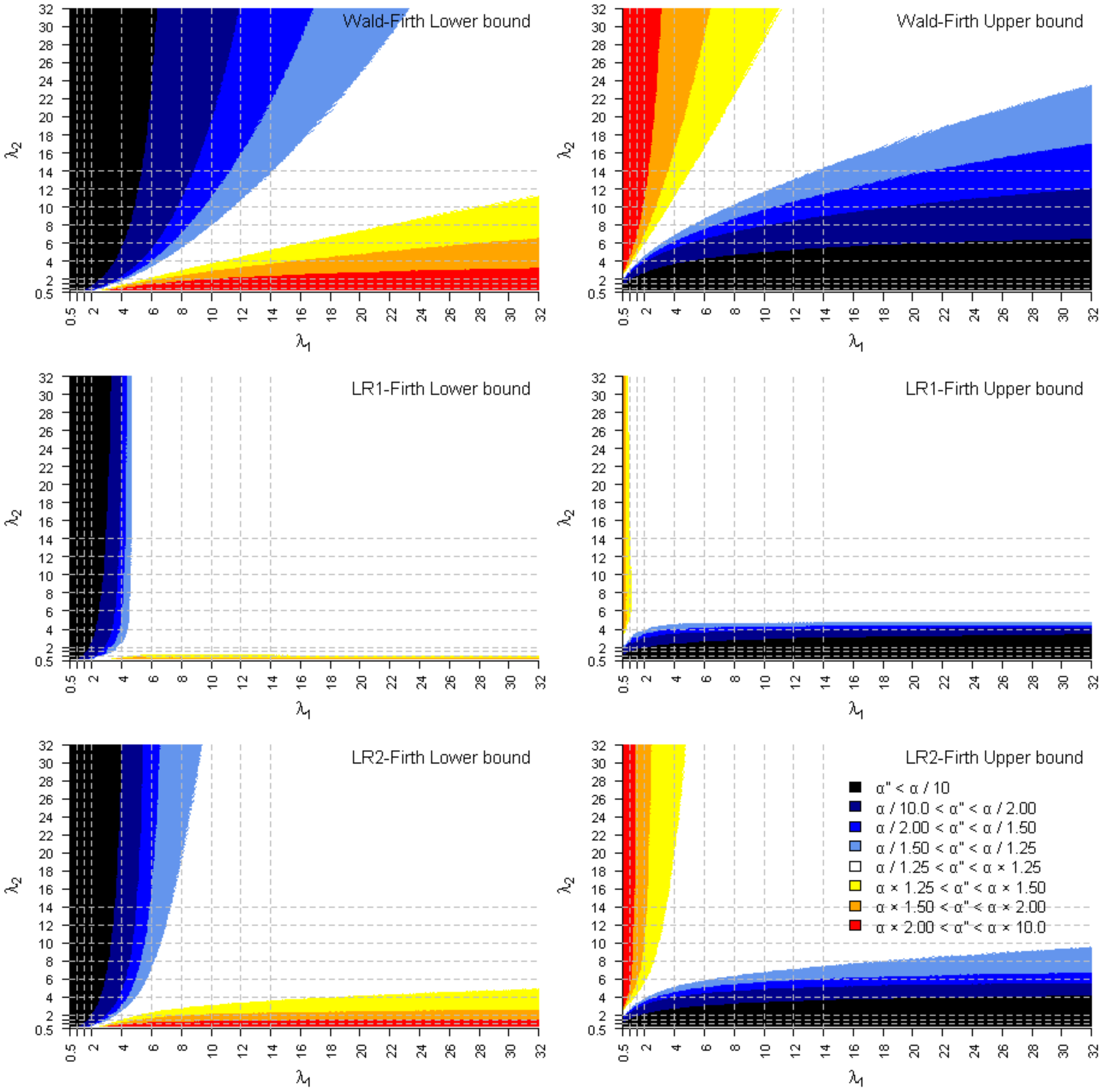

*Figure 2 : unconditional $\alpha_L''$ and $\alpha_U''$ risks for the 95% two-sided Wald-Firth, LR1-Firth and LR2-Firth confidence intervals of a ratio $\Lambda_2/\Lambda_1$ of two Poisson variables, according to $\lambda_1$ (x axis) and $\lambda_2$ (y axis) the expectancies of the Poisson distribution parameters assuming these parameters follow a log-normal distribution having a geometrical standard deviation equal to 1.1. The white, yellow and light blue zones show the zones of tolerated risk deflation/inflation by a factor less than 1.5.*

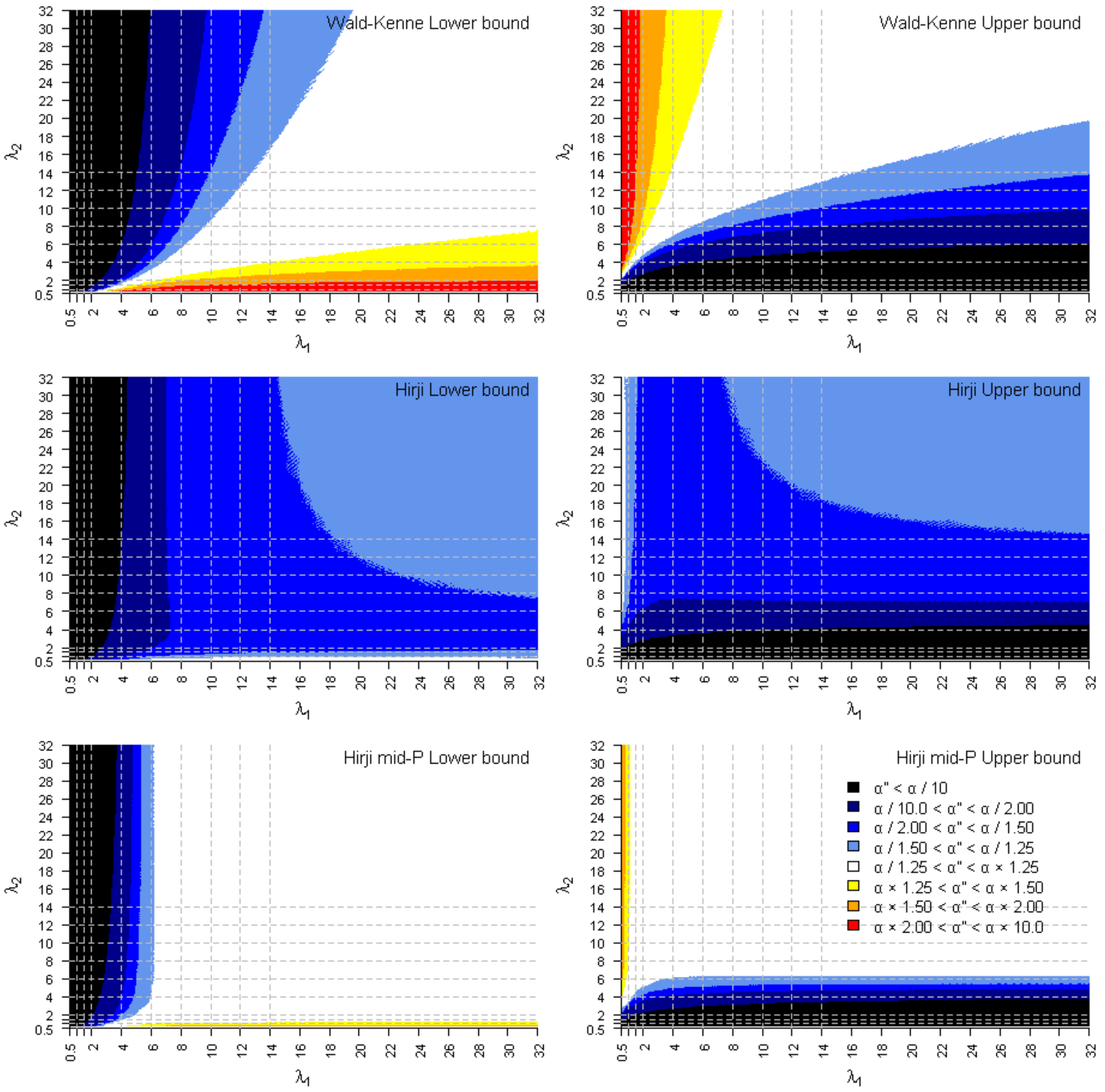

*Figure 3 : unconditional $\alpha_L''$ and $\alpha_U''$ risks for the 95% two-sided Wald-Kenne, Hirji and Hirji mid-P confidence intervals of a ratio $\Lambda_2/\Lambda_1$ of two Poisson variables, according to $\lambda_1$ (x axis) and $\lambda_2$ (y axis) the expectancies of the Poisson distribution parameters assuming these parameters follow a log-normal distribution having a geometrical standard deviation equal to 1.1. The white, yellow and light blue zones show the zones of tolerated risk deflation/inflation by a factor less than 1.5.*

Figure 1 shows that the ML LR CI estimator is much less biased than the Wald's and Score CI estimators, for both boundaries. The main bias of the ML LR CI estimator is due to an unavoidable over-coverage of the upper boundary when $\lambda_2$ (numerator) is close to zero and $\lambda_2/\lambda_1$ is small. Indeed, if the ratio $\lambda_2/\lambda_1$ is small enough to be smaller than the upper boundary of the CI for an observed denominator equal to 1, then, the upper boundary will never be below $\lambda_2/\lambda_1$ and the risk associated to the boundary will be zero. Similarly, a very small $\lambda_1$ and high $\lambda_2/\lambda_1$ ratio leads to an over-coverage of the lower boundary. There is some under-coverage for expected number of events ($\lambda_1$ or $\lambda_2$) smaller than 1. The score CI is not uniformly better than Wald's CI: it has less over-coverage but has more under-coverage.

Although Firth's estimator may be less biased than the ML estimator for a point estimate, Figure 2 shows that the associated Firth-Wald CI is more biased than Wald's CI associated to the ML estimator (Figure 1). The LR1-Firth CI is very slightly more conservative (more over-coverage, less under-coverage) than the ML LR CI (Figure 1). Approximation of the differences of penalized deviances to a chi-square distribution seems to lead to biased CI as seen in the LR2-Firth interval.

Figure 3 shows that the Kenne-Wald CI estimator is less biased (less over-coverage, less under-coverage) that the Firth-Wald CI estimator (Figure 2), but is still more biased than the standard Wald CI estimator (Figure 1). Hirji's CI, without mid-P modification, is strictly conservative, with much over-coverage and no under-coverage. The mid-P modification provides properties close to that of the ML LR CI estimator (Figure 1), but slightly more conservative.

## 5.2 Conditional coverage

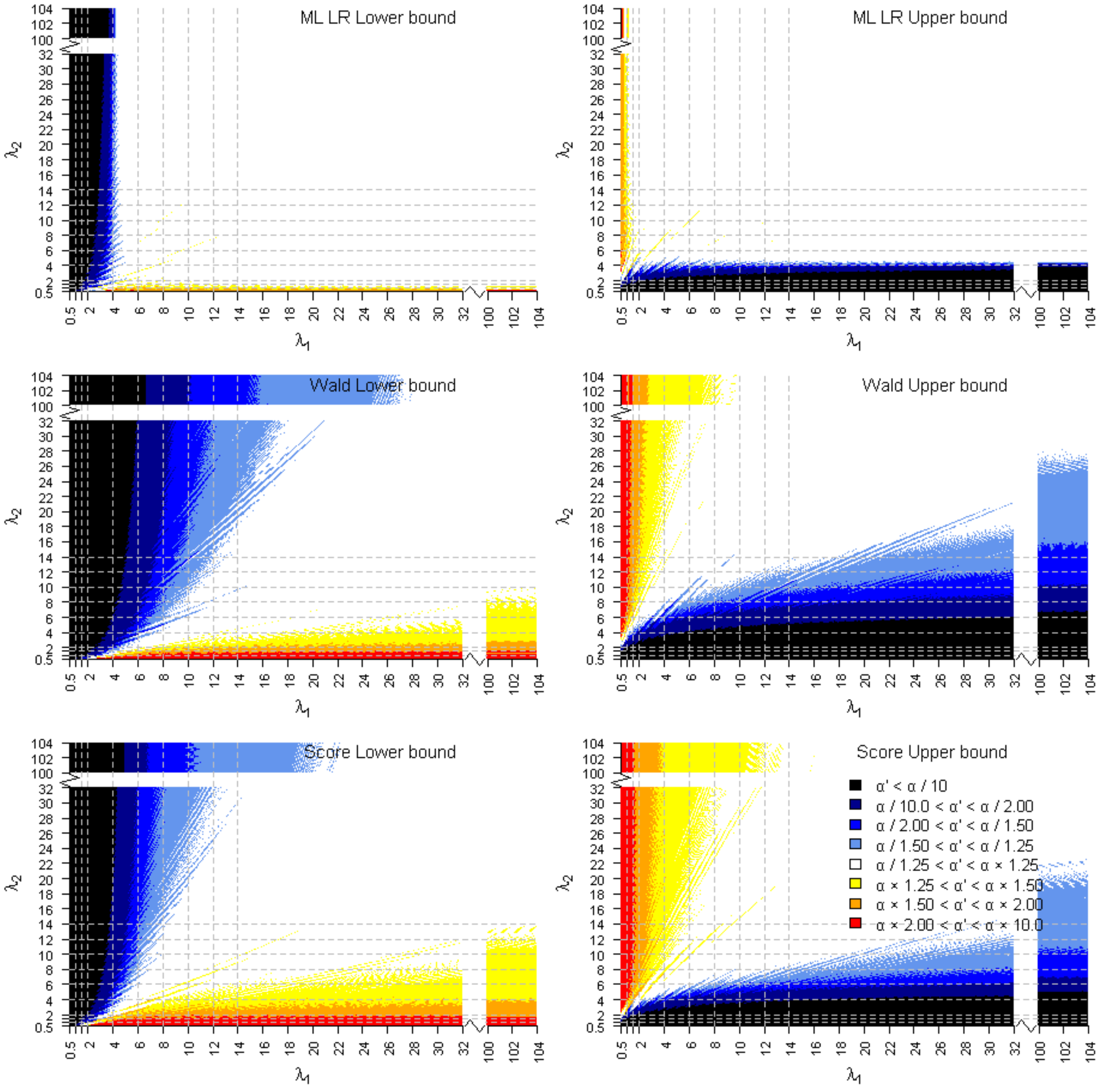

*Figure 4: conditional $\alpha'_L$ and $\alpha'_U$ risks for the 95% two-sided likelihood ratio confidence interval (LR), 95% two-sided Wald confidence interval and 95% two-sided Score confidence interval of a ratio $\lambda_2/\lambda_1$ of two Poisson variables, according to $\lambda_1$ (x axis) and $\lambda_2$ (y axis) the parameters of the Poisson distributions. The white, yellow and light blue zones show the zones of tolerated risk deflation/inflation by a factor less than 1.5.*

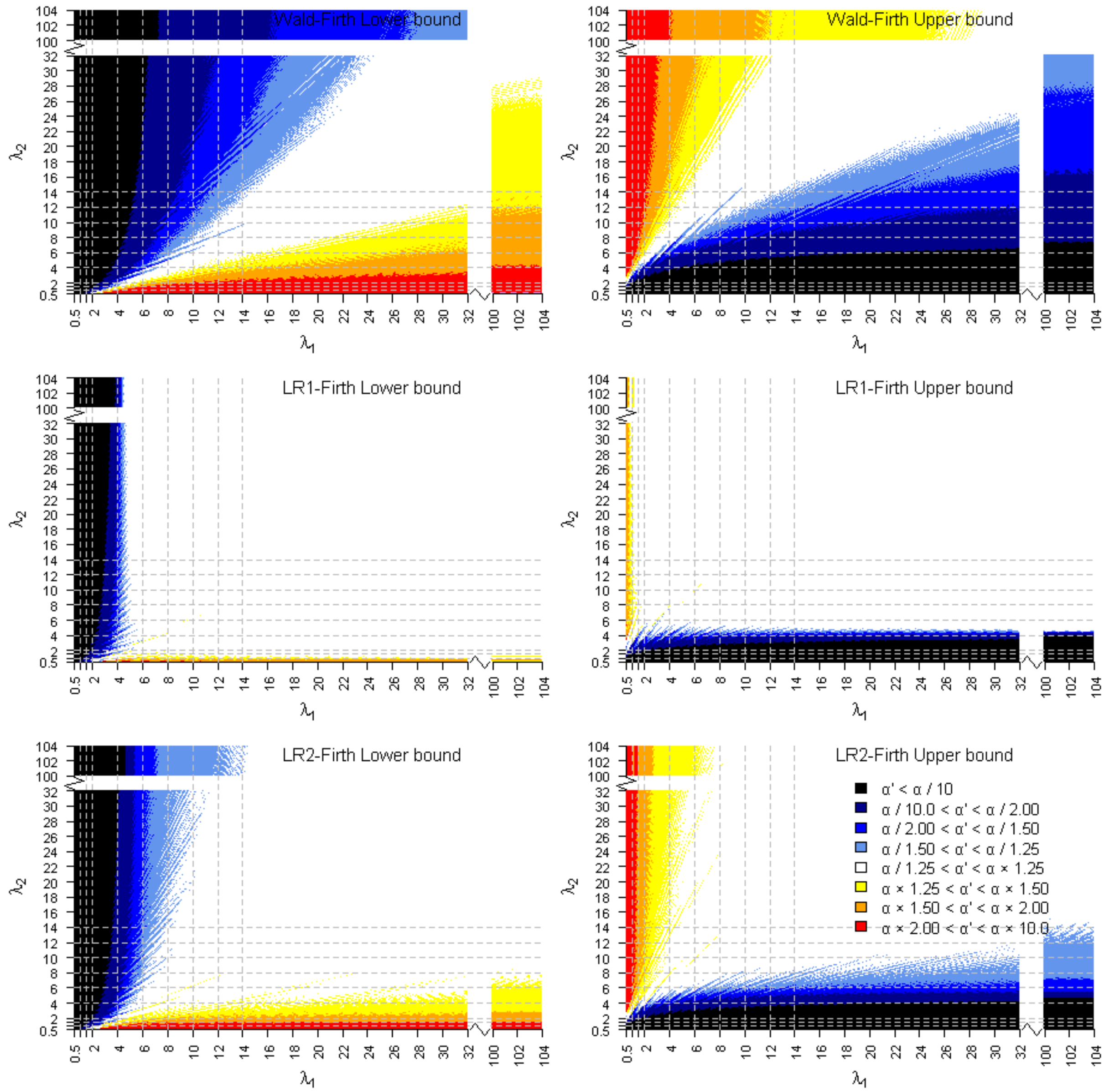

*Figure 5: conditional $\alpha'_L$ and $\alpha'_U$ risks for the 95% two-sided Wald-Firth, LR1-Firth and LR2-Firth confidence intervals of a ratio $\lambda_2/\lambda_1$ of two Poisson variables, according to $\lambda_1$ (x axis) and $\lambda_2$ (y axis) the parameters of the Poisson distributions. The white, yellow and light blue zones show the zones of tolerated risk deflation/inflation by a factor less than 1.5.*

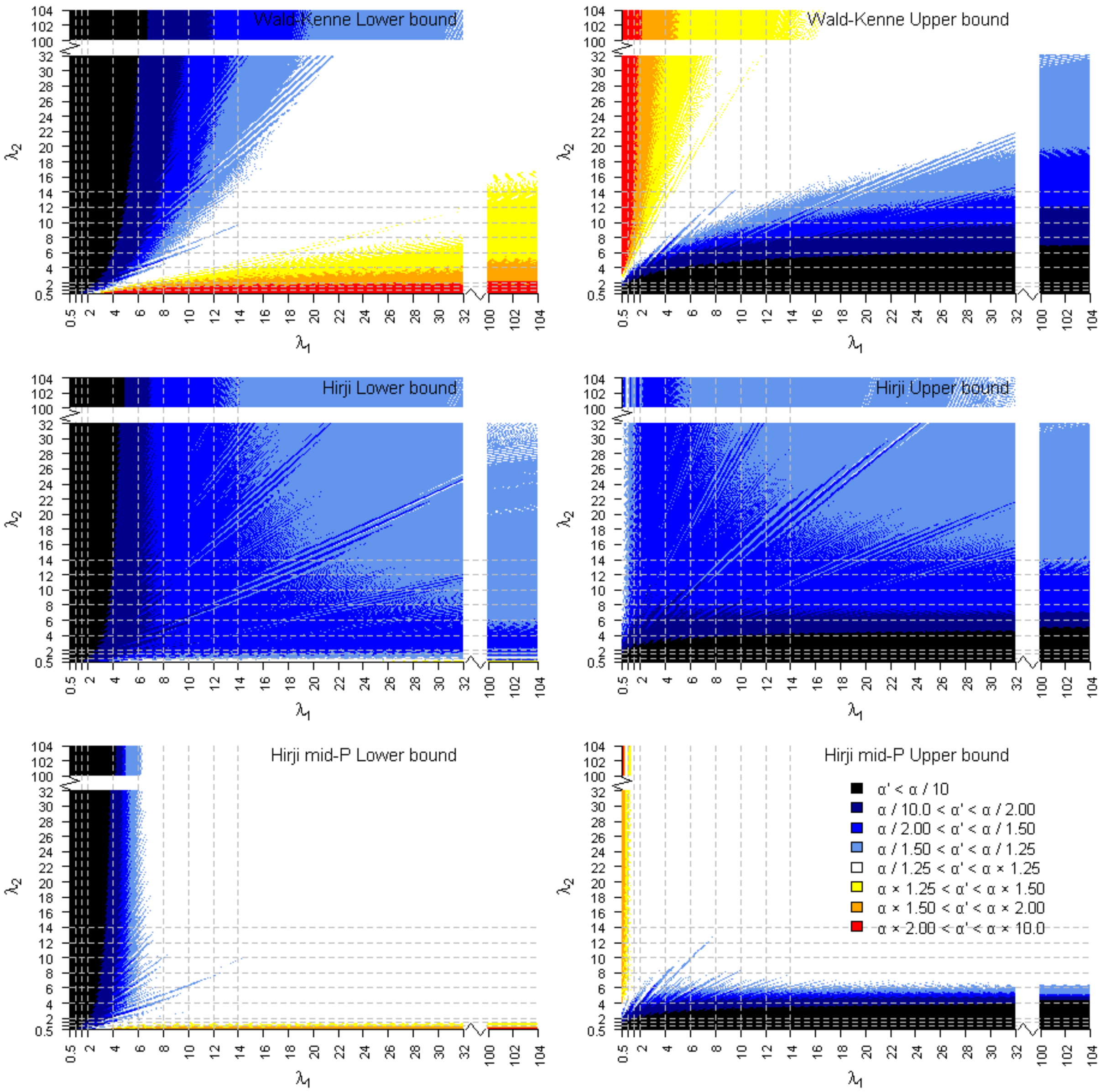

*Figure 6: conditional $\alpha'_L$ and $\alpha'_U$ risks for the 95% two-sided Wald-Kenne, Hirji and Hirji mid-P confidence intervals of a ratio $\lambda_2/\lambda_1$ of two Poisson variables, according to $\lambda_1$ (x axis) and $\lambda_2$ (y axis) the parameters of the Poisson distributions. The white, yellow and light blue zones show the zones of tolerated risk deflation/inflation by a factor less than 1.5.*

Figures 4 to 6 show that the plane of unconditional risks is not perfectly smooth as had been shown for the estimation of a single proportion by Brown, Cai and DasGupta [19], but otherwise confirm what was shown in figures 1 to 3. Oscillations are moderate because even if $\lambda_1$ and $\lambda_2$ are assumed to be constant, $y_1 + y_2$ is variable.

## 5.3 Interval half-widths

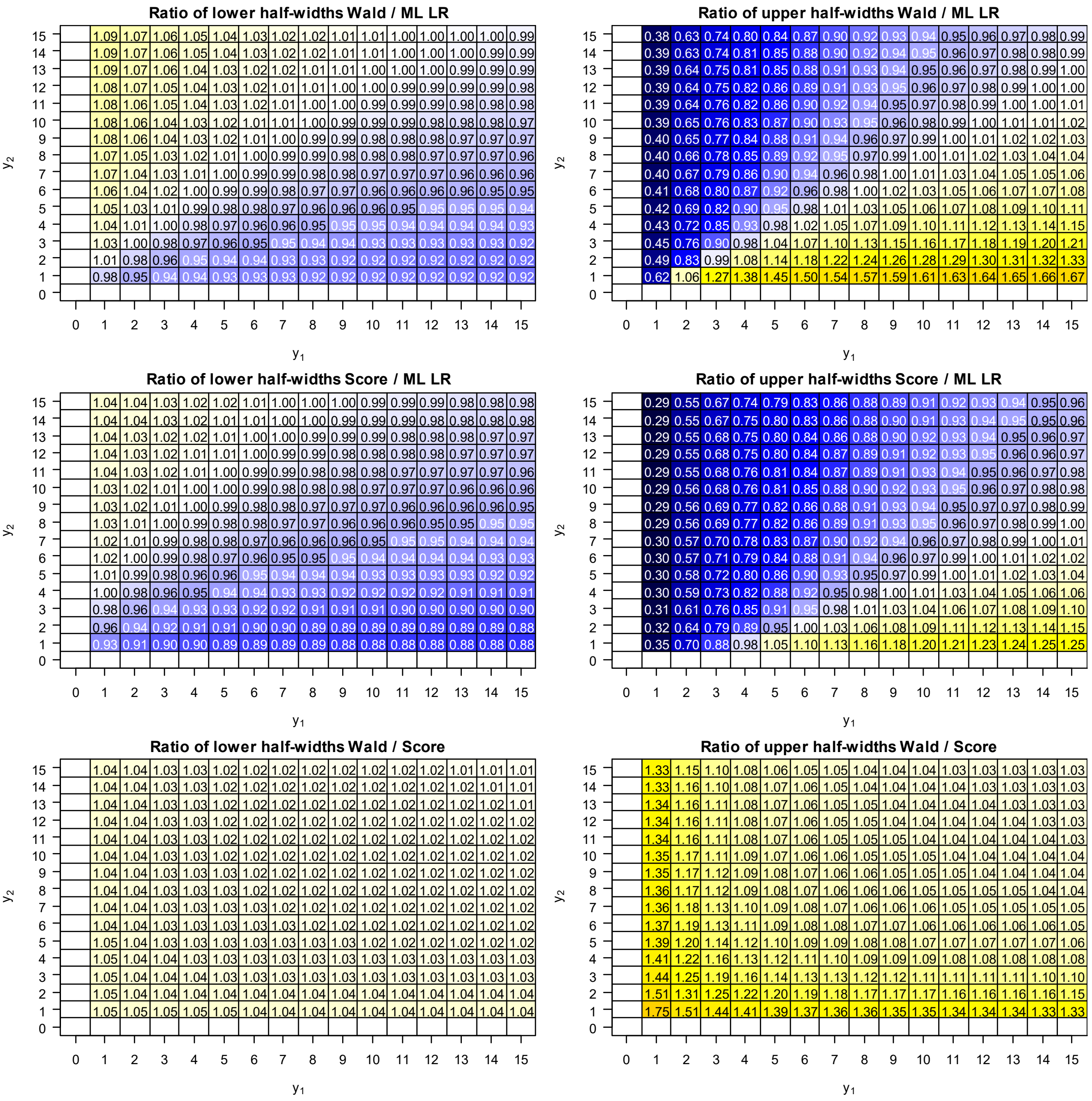

Figure 7: Ratio of half-widths (ML point estimate minus CI boundary) of Wald's, Score and ML LR 95% CI respectively to each other for $y_1$ (x axis) and $y_2$ (y axis), the observed number of events, from 1 to 15.

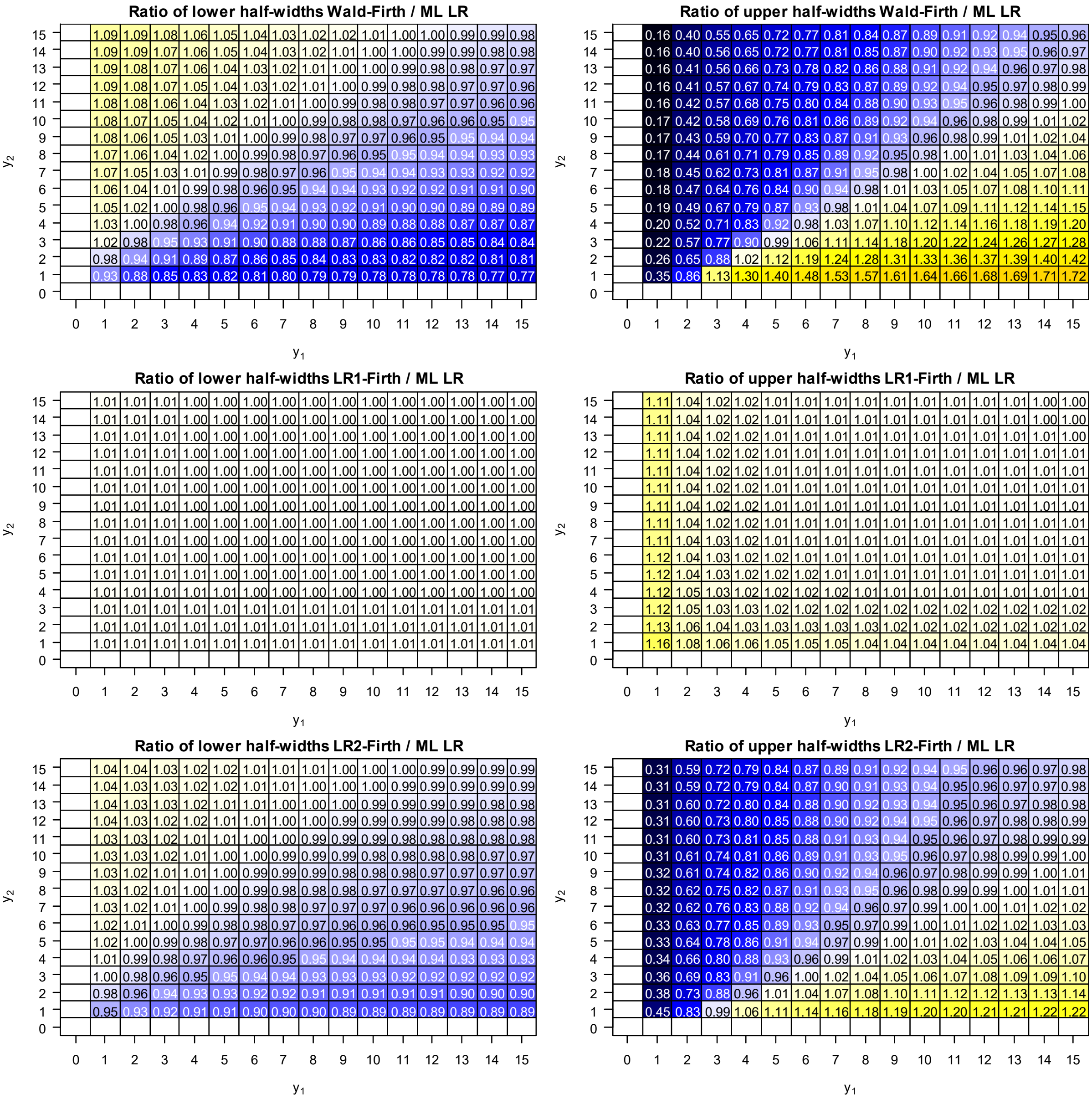

Figure 8: Ratio of half-widths (ML point estimate minus CI boundary) of the Wald-Firth, LR1-Firth and LR2-Firth 95% CI compared to the ML LR CI for $y_1$ (x axis) and $y_2$ (y axis), the observed number of events, from 1 to 15.

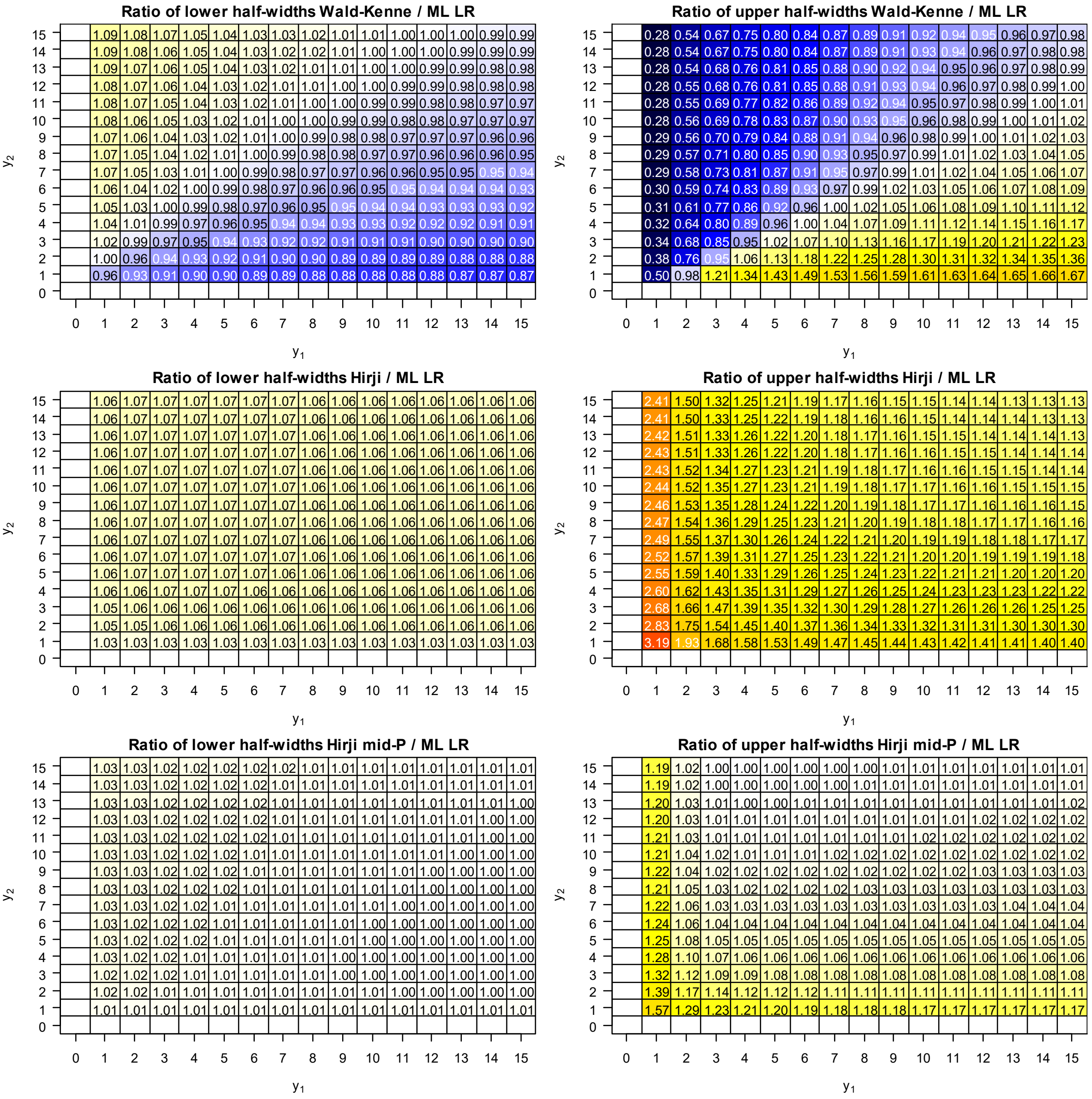

Figure 9: Ratio of half-widths (ML point estimate minus CI boundary) of the Wald-Kenne, Hirji and Hirji mid-P 95% CI compared to the ML LR CI for $y_1$ (x axis) and $y_2$ (y axis), the observed number of events, from 1 to 15.

Figures 7 to 10 show the half-widths of CI estimators relatively to the ML LR CI. Where a CI estimator is larger, it tends to over-cover and where it is shorter, it tends to under-cover.

## 5.4 Sensitivity analyzes: change in random distribution of $\Lambda_1$ and $\Lambda_2$ for unconditional risks

The log-normal distributions for $\Lambda_1$ and $\Lambda_2$ had a geometrical standard deviation equal to 1.10 in the primary analysis. Setting this geometrical standard deviation to 1.20 (respectively 1.05) changed, on average, the absolute coverage error of $1.4\times10^{-4}$ (resp $7.5\times10^{-5}$) on the whole set of points of Figures 1 to 3, with a 99th percentile equal to $1.0\times10^{-3}$ (resp $6.2\times10^{-4}$) and maximum equal to $7.7\times10^{-3}$ (resp 1.0%). The visual difference on the figures was negligible (not shown).

## 5.5 Secondary analyzes: change in confidence level

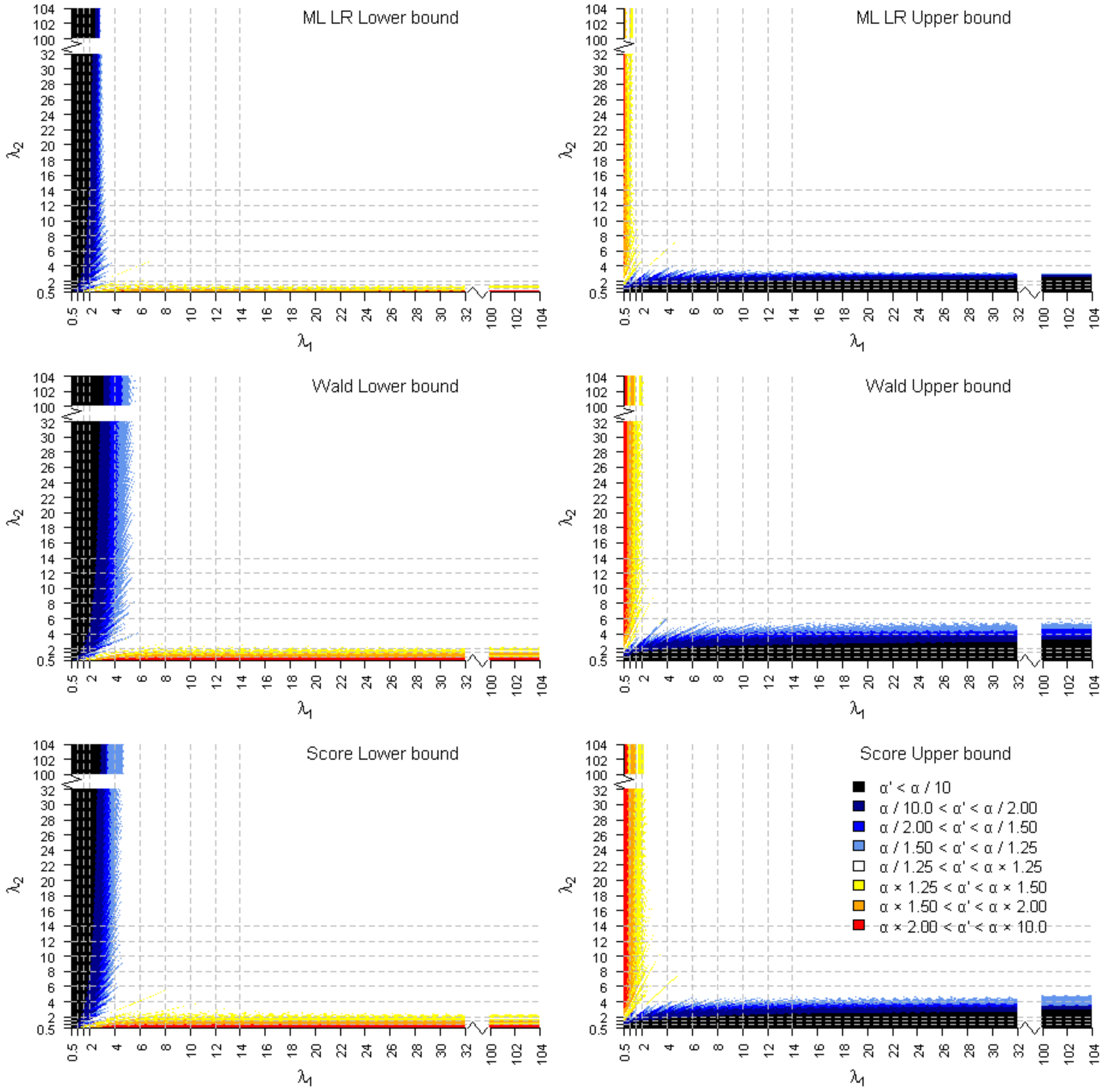

*Figure 10: conditional $\alpha'_L$ and $\alpha'_U$ risks for the 80% two-sided likelihood ratio confidence interval (LR), 80% two-sided Wald confidence interval and 80% two-sided Score confidence interval of a ratio $\lambda_2/\lambda_1$ of two Poisson variables, according to $\lambda_1$ (x axis) and $\lambda_2$ (y axis) the parameters of the Poisson distributions. The white, yellow and light blue zones show the zones of tolerated risk deflation/inflation by a factor less than 1.5.*

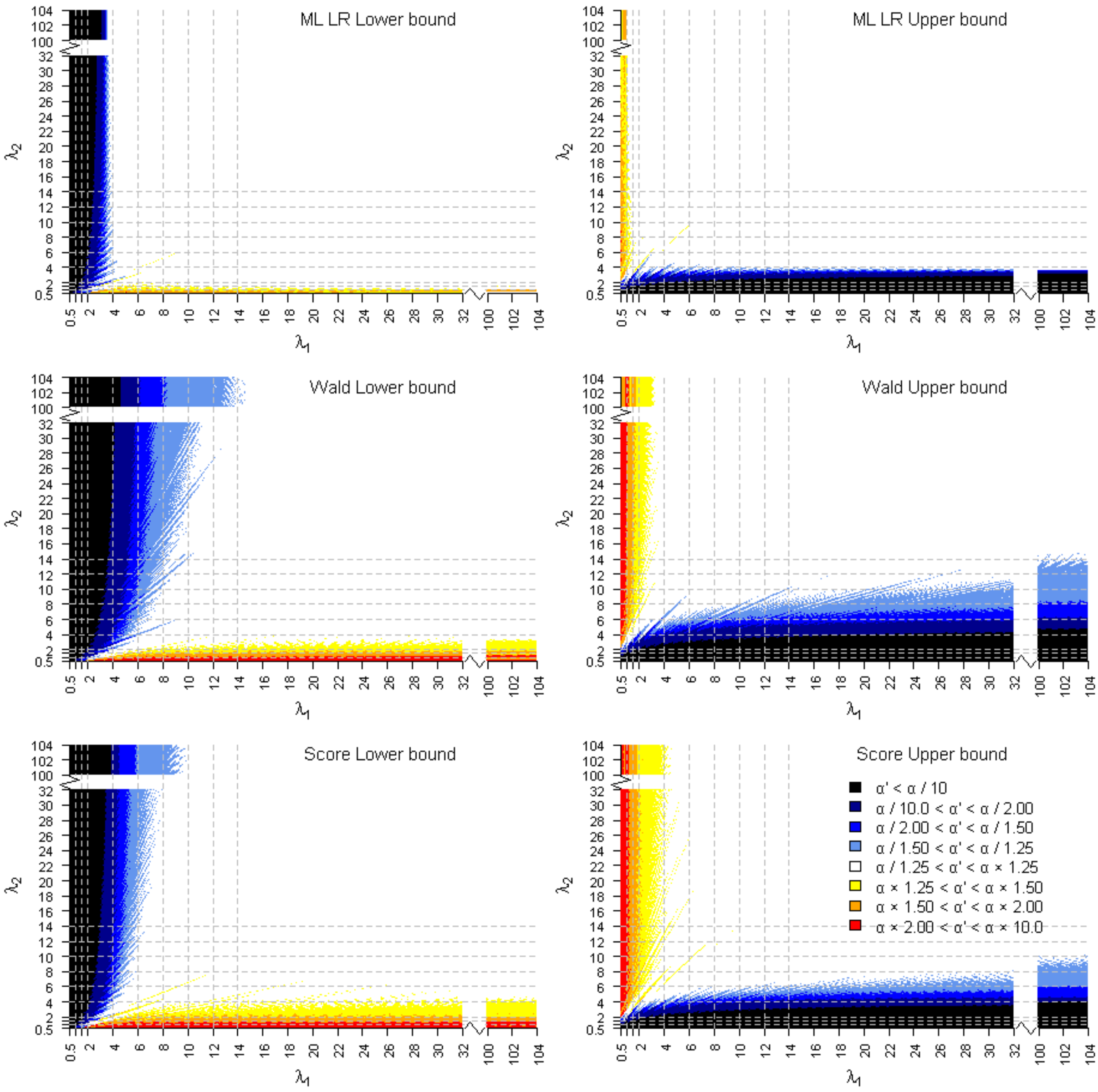

*Figure 11: conditional $\alpha'_L$ and $\alpha'_U$ risks for the 90% two-sided likelihood ratio confidence interval (LR), 90% two-sided Wald confidence interval and 90% two-sided Score confidence interval of a ratio $\lambda_2/\lambda_1$ of two Poisson variables, according to $\lambda_1$ (x axis) and $\lambda_2$ (y axis) the parameters of the Poisson distributions. The white, yellow and light blue zones show the zones of tolerated risk deflation/inflation by a factor less than 1.5.*

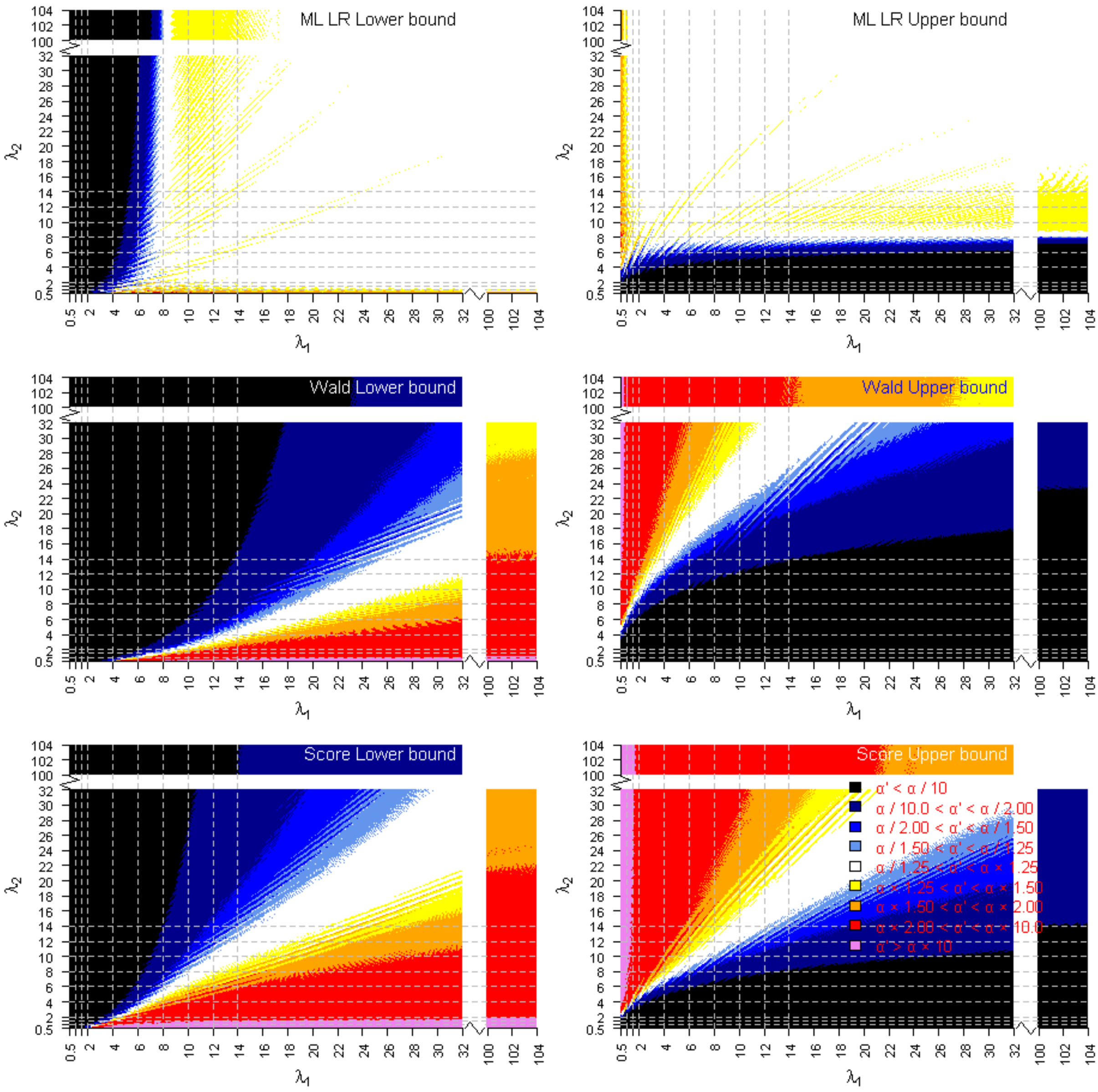

*Figure 12: conditional $\alpha'_L$ and $\alpha'_U$ risks for the 99.9% two-sided likelihood ratio confidence interval (LR), 99.9% two-sided Wald confidence interval and 99.9% two-sided Score confidence interval of a ratio $\lambda_2/\lambda_1$ of two Poisson variables, according to $\lambda_1$ (x axis) and $\lambda_2$ (y axis) the parameters of the Poisson distributions. The white, yellow and light blue zones show the zones of tolerated risk deflation/inflation by a factor less than 1.5.*

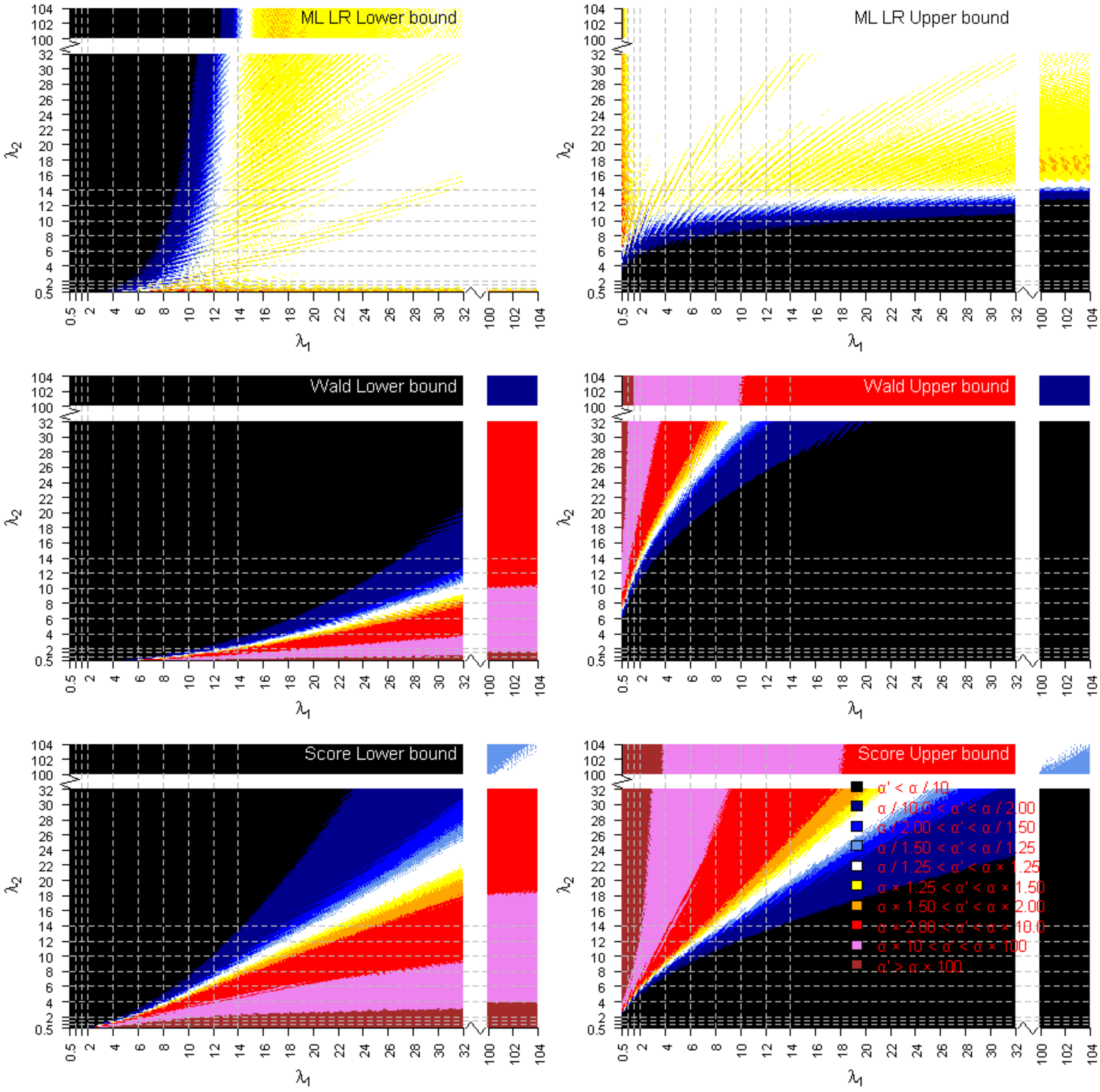

*Figure 13: conditional $\alpha_L'$ and $\alpha_U'$ risks for the 99.9999% two-sided likelihood ratio confidence interval (LR), 99.9999% two-sided Wald confidence interval and 99.9999% two-sided Score confidence interval of a ratio $\lambda_2/\lambda_1$ of two Poisson variables, according to $\lambda_1$ (x axis) and $\lambda_2$ (y axis) the parameters of the Poisson distributions. The white, yellow and light blue zones show the zones of tolerated risk deflation/inflation by a factor less than 1.5.*

Secondary analyzes show that coverage errors are worse for confidence levels close to 1. The ML LR CI behaves much better than Wald' and the Score CI for high confidence levels. Its main flaw is the unavoidable over-coverage when one or the other $\lambda_i$ is small. For $2 \times \alpha = 10^{-6}$ (cf Figure 13), the ML LR CI has some light to moderate under-coverage even for quite large $\lambda_1$ and $\lambda_2$ values (above 20).

When both $\lambda_i$ are large but that $\lambda_2/\lambda_1$ is far from 1, Wald's and the Score CI are biased. For instance, with $\lambda_1 = 100$ and $\lambda_2 = 400$ and $2 \times \alpha = 10^{-6}$, the conditional risk $\alpha_L' = \alpha \times 0.09$ and $\alpha_U' = \alpha \times 2.34$ for Wald's CI, $\alpha_L' = \alpha \times 0.18$ and $\alpha_U' = \alpha \times 2.91$ for the Score CI while $\alpha_L' = \alpha \times 1.08$ and $\alpha_U' = \alpha \times 0.96$ for the ML LR CI.

With $\lambda_1 = 100$ and $\lambda_2 = 400$ and $2 \times \alpha = 10^{-3}$, the conditional risk $\alpha'_L = \alpha \times 0.57$ and $\alpha'_U = \alpha \times 1.34$ for Wald's CI, $\alpha'_L = \alpha \times 0.63$ and $\alpha'_U = \alpha \times 1.41$ for the Score CI while $\alpha'_L = \alpha \times 1.05$ and $\alpha'_U = \alpha \times 0.97$ for the ML LR CI.

# 6 Discussion

## 6.1 Unconditional and conditional risks

Some "exact" CI estimators, such as Hirji's estimator [9], are designed to be strictly conservative. The actual coverage is always greater or equal to the nominal coverage. This estimator, like all other estimators described in this work, is conditional to the total number of events $y_1 + y_2$. For some values of $y_1 + y_2$ the actual coverage is close to 95% while it is closer to 97.5% for other values of $y_1 + y_2$. Hirji's estimator for the bivariate Poisson problem is equivalent to the Clopper-Pearson CI for a binomial distribution and the oscillations alongside $y_1 + y_2$ found by Brown Cai and DasGupta [19] apply to Hirji's estimator too. Since actually, $Y_1 + Y_2$ is random, oscillations are averaged and the actual coverage is always strictly greater than 95%. When analyzing the $\alpha'_L$ and $\alpha'_U$ risks with a random $Y_1 + Y_2$, Hirji's CI is too conservative (see Figure 6) leading to a wide interval (see Figure 9). As discussed in the rationale of the Materials & Methods section, not only $Y_1$ and $Y_2$ are random, but $\lambda_1$ and $\lambda_2$ should be considered random too, leading to even smoother unconditional coverage errors $\alpha''_L$ and $\alpha''_U$. The actual variance of the distributions of $\Lambda_1$ and $\Lambda_2$ is unknown but does not matter much as long as it is continuous, as shown in the sensitivity analysis.

Averaging risks is not a new idea in the assessment of CI estimator bias. For instance, Agresti and Coull had shown that "exact" CI estimators tended to be too conservative when the average risk was analyzed [20]. Nevertheless, they had averaged the risk over the whole space of theoretical values. They concluded that the score CI (Wilson's score CI) behaved well, while actually, undercoverage for some parameters may compensate over-coverage for other distant values, as we observed on Figure 1. Our unconditional coverage analysis performs local averaging only which is more relevant to a particular setting where parameters $\lambda_1$ and $\lambda_2$ are moderately variable.

## 6.2 What is the best CI estimator?

The score and Wald's CI showed coverage bias much higher than the ML LR CI or Hirji's estimator with mid-P. Penalized likelihood estimators (Firth's and Kenne's estimators) perform poorly with Wald's confidence boundaries. One may notice that the Wald-Firth estimator is equivalent to Wald's CI in a maximum likelihood model after adding 0.5 events in both groups (analysis not shown). The main advantage of penalized likelihood estimators over the ML estimator is the bias reduction of the point estimate in small samples and availability of an estimation when the number of events is zero in one group. The LR1-Firth CI behaves well but cannot be computed when zero events are observed in one group (complete separation). Indeed, the likelihood profile is monotone. The LR2-Firth CI can be computed in case of complete separation but has poor coverage properties. It is equivalent to a ML LR CI after adding 0.5 events in both groups (analysis not shown). Hirji's estimator is too conservative while Hirji's mid-P estimator is quite good but a bit wider and more conservative than the ML LR CI. The ML LR CI being theoretically simple, widely available and having very good coverage properties, we recommend its use in almost all scenarios.

## 6.3 When is the ML LR CI valid?

In the simple bivariate scenario, the ML LR CI always has tolerable coverage bias, for low or high confidence levels, as soon as the expected number of events is above 1 in both groups. When an experiment is performed, nobody knows the expected number of events $\lambda_1$ and $\lambda_2$. One can safely assume that both $\lambda_1 > 1$ and $\lambda_2 > 1$ if both $y_1 \geq 3$ and $y_2 \geq 3$. The highest risk of having both $y_i \geq$ 3 while one of the $\lambda_i$ is actually below or equal to 1 occurs when one of the $\lambda_i$ is high but the other is equal to 1. In that case, the risk of having both $y_i \geq 3$ is 8% according to the quantiles of the Poisson distribution $P(1)$. When one or both $y_i$ are between 1 and 3, we suggest to provide the ML LR CI if the estimation was planned, anyway. Not doing so would bias the literature towards overestimating the smaller $\lambda_i$ because the CI would not be provided when it is underestimated. We recommend that, in the first place, the estimation of $\lambda_2/\lambda_1$ be not attempted at all if there are prior information that suggests that either $\lambda_1$ or $\lambda_2$ is below 1. Indeed, the statistical precision will be unacceptable well before the statistical estimator is invalid. When $y_1 = 0$ or $y_2 = 0$ we assumed that the CI would not be provided so that all results of our analysis are conditional to both $y_i \geq 1$. Therefore, this selective reporting bias is taken in account in our results.

## 6.4 Unusual scenario where one $\lambda_i$ is large but not the other

Sometimes, one $\lambda_i$ may be expected to be large while the other is expected to be close to zero although the offsets (e.g. number of patients-years at risk) are not very different. This may happen when the outcome is expected to (almost) completely disappears in a group. In that case the estimation of the absolute risk difference in a linear model may be more relevant than the risk ratio and we suggest not to attempt estimation of the risk ratio in the first place. The second scenario occurs when the two groups have very different offsets (e.g. number of patients-years at risk) although the actual ratio assessed $\frac{\lambda_2/r_2}{\lambda_1/r_1}$ may not be very far from 1. In that case one of the $\lambda_i$ may be close to zero and the estimation of the ratio almost impossible due to poor statistical precision. In that case, we advise to change the design of the study (e.g. cohort -> case-control). In all other cases, both $\lambda_i$ are expected to be large enough to warrant good coverage to the ML LR CI.

## 6.5 Hypothesis tests

Since all CI estimators analyzed can be obtained by inversion of hypothesis tests, our results apply to hypothesis tests as well. As expected, approximations are best for large P-values, and bias may be enormous in a scenario where the significance level is very close to zero (see Figure 12 and Figure 13). This may happen in a multiple-testing scenario with thousands of tests. In that case, the Wald and score tests are very biased, even when the number of events is quite large ($\lambda_1 = 100$ and $\lambda_2 = 400$). Due to different offsets in both groups, the ratio actually tested may be equal to 1 but the offset (e.g. number of patients-years at risk) $r_2$ may be four times larger than $r_1$ so the $\lambda_2 = 4 \times \lambda_1$ too. Even outside of the multiple testing scenario, Wald's and the Score CI may provide a wrong impression of confidence when the P-value is very small.

## 6.6 Software implementation

By default R statistical software uses the ML LR CI for generalized linear models with the function confint but the function summary on GLMs uses Wald's method for hypothesis tests. The package glmglrt, created by the author of this article and available on the Comprehensive R archive Network (CRAN), adds a column “LR P value” to the summary.glm function without changing other columns, for backwards compatibility.This function is compatible with most features of GLM (weights, offsets, dropped coefficients). When convergence in the null hypothesis is impossible (quite frequent for Intercept of log-binomial models), it provides "NA" in place of the P-value.

SAS (Statistical Analysis System Institute, Inc., Cary, NC) software provides options "LRCI" and "type3" to the MODEL statement in the GENMOD procedure, providing respectively ML LR CI and LR tests. Unfortunately, if a categorical variable is included in a model, it's automatically recoded as several binary variables but the "type3" statement tests the hypothesis that all coefficients of the variable are equal to zero at once. Manual recoding of categorical variables to binary variables is required.

Stata (StataCorp LP, TX, USA) provides the command `pllf` [21] to compute profile-likelihood CIs in various models and LR hypothesis tests can be performed by the `lrtest` command; the latter requires fitting two models and saving estimates with the `estimates store` command.

SPSS does not provide any easy way of computing ML LR CI or performing LR tests.

### 6.7 Limits of this work

The case where one $y_i$ is zero (complete separation) has been excluded from all analyzes. Therefore, our results apply to the case where the statistician do not provide any CI in that case. These results do not apply to a statistician who prefer to provide Hirji's exact CI or a penalized likelihood CI when one of the $y_i$ is zero. As discussed above, the study should be designed to avoid these cases in the first place.

The logistic regression has not been analyzed. The Poisson regression has been considered as the limit case of the logistic regression, but actually, the logistic regression may behave quite differently when proportions are close to 50%. Since the bias of Wald's CI is mainly due to the skewness of the ML estimator (data not shown), the case where both proportions are close to 50% may be optimal, while the case where one proportion is close to 0 or 100% and the other is on the opposite side of the scale, should be the worst.

The effect of additional covariates (multivariate regression) has not been analyzed. The problems found in the simple bivariate scenario are expected to be found in multivariate regressions, but additional problems, such as inflation of effects due to overfitting are expected with the ML LR CI. The overfitting problem should be slightly weaker with penalized likelihood CIs. One may note that penalized likelihood estimation is not the same as penalized regression (LASSO, Ridge Regression, Elastic-Net). The former avoids point estimate bias but should have a very small effect on over-fitting.

Half-widths of CI estimators can hardly be compared when their coverage bias is not identical. Computing half-widths after correction of the coverage bias by inflation/deflation of the confidence level was attempted. A global correction would not make sense as it could not correct local coverage errors. A local correction made more sense but would transform so much the CI estimator that all CI estimators would become almost identical, with half-width differences that would be due to the method of correction rather than to real differences. Fortunately, for complex multivariate statistics, some estimators may have much better performances (lower variance and same bias) than others, but for statistics as simple as a rate ratio, that does not seem to be the case.

## 7 Conclusion

From theoretical considerations and results of this work, we advise not to attempt assessment of an incidence ratio when the frequency in one group is expected to be less than 1 on average. Otherwise we advise the LR hypothesis test be performed and the ML LR CI be computed for the rate ratio estimated in Poisson regressions with log link as long as there is no group where the number of events is zero (complete separation). In case of complete separation, one may not publish the rate ratio CI (strategy assessed in this work), or Firth's estimator may be considered (not assessed in this work).